# Magnetoresistance of layered conductors under conditions of topological phase transition

O.Galbova[1)], V.Peschansky[2,3)]

[1)]Faculty of Natural Sciences and Mathematics, Institute of Physics, P.O.Box 162, 1000, Skopje, Republic of Macedonia

[2)]B.I.Verkin Institute for Low Temperature Physics and Engineering, National Academy of Sciences of the Ukraine, Kharkov, 61103, Ukraine

[3)]National University "V.N.Karazina" of Kharkov, Kharkov-61077, Ukraine

**Abstract**

The resistance of layered conductors with a multysheet Fermi surface (FS), in a high magnetic field, in the immediate vicinity of Lifshic's topological transition when the separate FS sheets are drown together by an external action, pressure in part (and eventual change of the FS connectivity) is studied theoreticaly. Analysis of magnetoresistance near topological transition is illustrated for the case of FS in the shape of lightly corrugated cylinder and two corrugated planes distributed with a repeated period in the pulse space. It yields, that as the FS plane sheets approach sufficiently the cylinder, the charge carriers produce a magnetic breakdown of one FS sheet to another, decreasing a sharp anisotropy of magnetoresistance to the in-plane current. Instead of square increase with a magnetic field, the slower resistance growth remains linear in the field within a broad magnetic-field range. In the intimate vicinity of topological transition, when the energy gap between FS layers is negligibly small, the resistance is saturated.

_______________________________________________________________________

The electronic phenomena that occur in conductors in the presence of strong magnetic fields are highly dependent on the energy spectrum of conduction electrons (responsible for the transfer of charges). In crystal structures the periodic dependence of the quasiparticles' energy $\varepsilon(\vec{p})$ on the momentum, $\vec{p}$, is essentially different from that of free electrons. The latter is at the origin of a number of physical phenomena. The experimental study of these phenomena allows gaining thorough information about the energy spectrum of the solids.

The investigation of the dependence of the resistance (in metals) on the magnitude and the orientation of the magnetic field, with respect to a crystal axis of the sample, allows for complete determination of the topology of the Fermi surface (FS) $\varepsilon(p) = \varepsilon_F$ – a basic/fundamental characteristic of the electron energy spectrum [1]. To accomplish the latter it would be enough to have perfect single crystal samples with large mean free path length, $l$, of the charge carriers. In such a case, namely, during a time period equal to $\tau = l/v$ ($\tau$ is the relaxation time) the carriers perform several revolutions in a magnetic field given with a circular frequency $\omega_c = eH/(m^*c)$ where $c$ is the speed of light, and $e$, $v$, $m^*$, and $\varepsilon_F$ are the charge, velocity, the cyclotron effective mass and the Fermy energy, respectively).

The investigations of the quantum oscillation effects of Shubnikov–de Haas [2] and of de Haas–van Alphen [3], as well as a number of high frequency phenomena in strong magnetic fields [4,5] allow for a complete determination of the form of FS.

The inverse problem, namely calculation of the electronic energy spectrum in metals from experimental data, proved to be really promising. This trend in the electronic theory of metals was later called 'Fermiology'.

In the last few decades the methods of Fermiology were successfully applied in different structures containing charge carriers (mainly layered structures and organic string-like structures exhibiting strong conductance anisotropy).

The interest in the above emerged since the discovery of the new superconductors characterized with large values for the critical parameters. According to Little [6] it is expected that exactly the low dimensional conductors will show high-temperature transitions in their superconductor phase. As a consequence, a number of low dimensional structures with unique properties were synthesized.

A significant part of the layered conductors show metallic conductivity not only in the plane of the layer, but also along the normal to the layer $\vec{n}$. A strong anisotropy in the electrical conductivity of the layered conductors is probably related to the weak dependence of the energy of the charge carriers

$$\varepsilon(p) = \sum_{n=0}^{\infty} \varepsilon_n(p_x, p_y) \cos\left(\frac{anp_z}{\hbar} + \alpha_n(p_x, p_y)\right) \quad (1)$$

on the $z$-component od the momentum, $p_z = \vec{p} \cdot \vec{n}$, so that their velocity along the normal to the layer

$$\varepsilon_n(-p_x, -p_y) = \varepsilon_n(p_x, p_y) \quad ; \quad \alpha_n(-p_x, -p_y) = \alpha_n(p_x, p_y) \quad (2)$$

is considerably smaller compared to the Fermi velocity $v_F$ with which the electrons move within the layer.

In the above equation, $a$ is the separation between the layers, while $\varepsilon_n(p_x, p_y)$ are arbitrary functions. The quasi-two-dimensionality parameter of the electronic energy spectrum, $\eta$, is defined as a ratio between the maximum value for the projection of the velocity, $v_z$, on the Fermi surface and the Fermi velocity, $v_F$.

The up-to-date techniques for preparation of organic single crystal conductors allow the precondition $\omega_c \tau \ll 1$ to be fulfilled. The latter is a necessary precondition for the Fermiology methods to be used in the obtainable magnetic fields.

The Fermi surface of the layered conductors is open and weakly corrugated along the $p_z$-axis. It can also be in a form of multysheet, or can consist of topologically different elements (e.g. weakly corrugated cylinders or planes), in the momentum space.

The experimental observations performed in the laboratory Shegoleva (Chernogolovka) on pronounced Shubnikov–de Haas effects, in magnetic field strengths of up to 14 T, stimulated further intense studies of electronic phenomena in low-dimensional structures with charge carriers. The sample was an organic layered conductor $\beta$-(BEDT-TTF)$_2$IBr$_2$. Under such strong fields, the behaviour of the magnetic resistance normal to the layer, on the field orientation, is highly unusual [8].

The dependence of the magnetic resistance on the angle $\theta$ between the $\vec{H}$ vector and the normal $\vec{n}$ to the layer exhibits sharp maxima, the positions of which repeat periodically as $\tan\theta$ functions for a wide interval of $\theta$ values. The periods of these 'angular' oscillations carry important informations about the 'size' of the FS [9], or the extent to which the FS sheets are corrugated [10]. The experimental study of this oscillatory effect at different orientations of the magnetic field with respect to a crystal axis of the sample allows complete determination of the FS shape, without using of other spectroscopic methods. The solution of the inverse problem for gaining knowledge about the FS on the basis of experimental data in layered structures was particularly successful in layered structures of the type of tetrathiafulvalene [11–20].

Let us consider layered conductors with a multysheet FS, under conditions of phase topological transition of Lifshic type [21]. As a result of an external generalized force (e.g. the pressure), the sheets of the FS come closer, that lead change of the FS connectivity. In the vicinity of the topological transition, when the FS sheets appear to be so close that the charge carriers (as a result of magnetic breakdown) can move between the sheets, their motion in the magnetic field becomes complex and unsolvable.

Let the Fermi surface be composed of a weakly corrugated cylinder and two corrugated planes, periodically repeats in the momentum space. To simplify things, let the $p_x$ axis to be normal to the sheet.

The current density, in the $\tau$-approximation for the collision integral, takes the form:

$$J_i = \sigma_{ij} E_j = -\frac{2e^2 H}{c(2\pi\hbar)^3} \int d\varepsilon \frac{\partial f_0(\varepsilon)}{\partial \varepsilon} \int dp_H \int dt \cdot v_i(t, p_H) \psi(t, p_H) = \langle v_i \psi \rangle \qquad (3)$$

where $f_0(\varepsilon)$ – is the equilibrium Fermi function for distribution of the charge carriers, $t$ – is the time they spent moving in magnetic field along trajectories given by $p_H$ = const, $\varepsilon$ = const, and the function $\psi(t, p_H)$ is given by

$$\psi(t, p_H) = \int_{\lambda_1}^{t} eEv(t') \exp\frac{t'-t}{\tau} dt + \psi(\lambda_1, p_H) \exp\frac{\lambda_1 - t}{\tau} \qquad (4)$$

The first term on the left-hand side of (4) gives the energy that conduction electrons gain in electric field $E$, while the function

$$\psi(\lambda_1, p_H) = \int_{-\infty}^{\lambda_1} eEv(t) \exp\frac{t - \lambda_1}{\tau} dt \qquad (5)$$

describes the complex motion of the electrons along the magnetic breakdown trajectories, with a probability for a magnetic breakdown in regions A and B in the moments $\lambda_1, \lambda_2, \lambda_3$ ($\lambda_1$ closest to the moment when electrons move from a given sheet of FS to a neighbouring one, and also $\lambda_j > \lambda_{j+1}$).

As several groups of charge carriers exist, each of them contributes to the current density

$$\langle v\psi \rangle = \langle v^{(1)}\psi \rangle + \langle v^{(2)}\psi \rangle + \langle v^{(3)}\psi \rangle \qquad (6)$$

where $\langle v^{(2)}\psi \rangle$ is the contribution to the current from states that, in the moment $t$, belong to the weakly corrugated cylinder, while the rest of the terms in (6) give the contribution to the current from the other electrons, with states belonging to sheets of the FS.

The very existence of FS sheets results in strong anisotropy of the magnetic resistance in a strong magnetic field, even in the (layers of the sample [22]. Providing that the probability $w$ for magnetic breakdown is negligibly small, the component $\sigma_{xx}$ of the conductivity tensor is comparable with the conductivity $\sigma_0$ in the absence of magnetic field. This is related to the existence of open trajectories of the charges in the FS sheets, that drift down the $x$-axis with a mean velocity of $\bar{v}_x$. In a magnetic field $\vec{H} = (0, H\sin\theta, H\cos\theta)$, the charge carriers may also drift along the $y$-axis with a velocity $\bar{v}_y = \bar{v}_z \tan\theta$ (the latter being significantly smaller than $\bar{v}_x$ due to $\tan\theta < 1$). Under these circumstances, the resistance in the plane of the layer increases with increasing the magnetic field strength ($\sim H^2$), showing a deep minimum for the field strength at which the current starts to flow in the $x$ direction. When the FS sheets approach the weakly corrugated cylinder, the probability for magnetic breakdown increases essentially, thus the electrons, starting from a FS sheet, begin to move along magnetic breakdown trajectories. In this way, their acceleration in an electric field (in the $x$-direction) is suppressed, because the velocity $v_x$ on the opposite FS sheet is of opposite sign.

As a consequence, an increase in the probability $w$ decreases the $\sigma_{xx}$ component of the conductivity, resulting in essential change of the dependence of the current resistance in the layers, on the magnetic field strength. Let us assume that the quasi-two-dimensional-parameter, $\eta$, of the electronic energy spectrum, is the smallest parameter in the problem. Further we will ignore the effect the weak corrugation of the FS along the $p_z$ direction might have on the probability $w$. Under the above assumptions, the motion along the magnetic breakdown trajectories is periodic, and the quantities $\lambda_j$ and $\lambda_{j+1}$ correspond to the period $T$ of a FS plane sheet, i.e. to a half-period $T'$ of a cylindrical FS sheet.

The function $\phi_i$, before and after the magnetic breakdown, satisfies the relation

$$\phi_i(\lambda_j + 0) = (1-w)\phi_i(\lambda_j - 0) + w\phi_{i+1}(\lambda_j - 0) \quad I=1,2,3 \ . \tag{7}$$

The functions $\phi_i(\lambda_j - 0)$ before, and $\phi_i(\lambda_{j+1} + 0)$ after the magnetic breakdown, in the moment $\lambda_{j+1}$ are related by

$$\phi_i(\lambda_j - 0) = A_i + \exp\left(\frac{\lambda_{j+1} - \lambda_j}{\tau}\right)\phi_i(\lambda_{j+1} + 0) \tag{8}$$

where

$$A_i = \int_{\lambda_{j+1}}^{\lambda_j} dt' e v^{(i)}(t', p_B) \mathrm{E} \exp\left(\frac{t' - \lambda_j}{\tau}\right) \tag{9}$$

is the energy produced in the electric field, between two instants of magnetic breakdown during the motion of electrons in the i-th sheet of the FS.

Using relations (7) и (8) one can easily find the relation between the functions $\phi_i(\lambda_j + 0)$ и $\phi_i(\lambda_{j+2} + 0)$. For the charge carriers starting their drifting from the first sheet of the FS one could write

$$\phi_1(\lambda_1 + 0) = (1-w)(A_1 + \exp(-T/\tau)\phi_1(\lambda_2 + 0)) + w(A_2 + \exp(-T'/\tau)\phi_2(\lambda_2 + 0)) \qquad (10)$$

and, quite analogously

$$\phi_2(\lambda_2 + 0) = (1-w)(A_2' + \exp(-T'/\tau)\phi_2'(\lambda_3 + 0)) + w(A_3 + \exp(-T/\tau)\phi_3(\lambda_3 + 0)) \qquad (11)$$
$$\phi_3(\lambda_3 + 0) = (1-w)(A_3 + \exp(-T/\tau)\phi_3(\lambda_4 + 0)) + w(A_2' + \exp(-T'/\tau)\phi_2'(\lambda_4 + 0)) \qquad (12)$$
$$\phi_2'(\lambda_4 + 0) = (1-w)(A_2 + \exp(-T'/\tau)\phi_2(\lambda_5 + 0)) + w(A_1 + \exp(-T/\tau)\phi_1(\lambda_5 + 0)) \qquad (13)$$
$$\phi_1(\lambda_5 + 0) = (1-w)(A_1 + \exp(-T/\tau)\phi_1(\lambda_6 + 0)) + w(A_2 + \exp(-T'/\tau)\phi_2(\lambda_6 + 0)). \qquad (14)$$

$A_2$ is the energy gained by the electrons in electric field, during their motion along the lower arc of the corrugated cylinder's section, and $A_2'$ refers to the upper arc (cf. Figure 1). In the zeroth approximation of the small parameter $\eta$, it follows that $A_2' = -A_2 \exp(-T'/\tau)$, and $A_3 = -A_1$.

Figure 1.

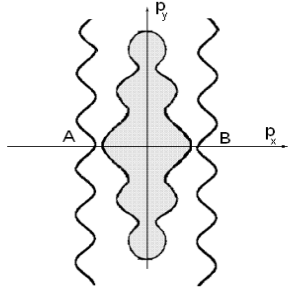

Applying these recursive relations several times, one could derive relations for $\phi_1(\lambda_1 + 0)$, $\phi_2(\lambda_1 + 0)$ and $\phi_3(\lambda_1 + 0)$ as sums of terms, proportional to the energy gained in the electric field between two instants of possible magnetic breakdowns. It is easy to notice that the recursive relations 11–14 repeat periodically, forming geometric progression with a quotient $w^4 \exp(-2\dfrac{T+T'}{\tau})$, thus enabling summation of the fast converging series

$$\phi_1(\lambda_1 + 0) = wA_2 + w^2 \exp(-T'/\tau)A_3 + w^3 \exp(-(T+T')/\tau)A'_2 + w^4 \exp(-(T+2T')/\tau)A_1 + ...$$
$$+ (1-w)\{(A_1 + \exp(-T/\tau)\phi_1(\lambda_2 + 0)) + w\exp(-T'/\tau)(A'_2 + \exp(-T'/\tau)\phi'_2(\lambda_2 + 0)) +$$
$$+ w^2 \exp[-(T+T')/\tau](A_3 + \exp(-T/\tau)\phi_3(\lambda_4 + 0) +$$
$$w^3 \exp[-(T+2T')/\tau](A_2 + \exp(-T'/\tau)\phi_2(\lambda_4 + 0) + ...\}$$
(15)

We will skip the derivation of the very complex algebraic expressions for the components of the conductivity tensor, and turn our attention to the analysis of some important limiting cases.

When the FS sheets come very close to each other, and the energy gap, $\Delta$, between them is negligibly small, the probability for a magnetic breakdown $w = \exp(-\Delta^2/\hbar\omega_c\varepsilon_F)$ is close to 1. Then, since the quantity $(1-w) \ll 1$, one gets

$$\phi_1(\lambda_1 + 0) = \frac{-w^3 A_1 \exp[-(T+T')/\tau]\{1 - w^2 \exp[-(T+T')/\tau]\}}{1 - w^4 \exp[-2(T+T')/\tau]} +$$
$$+ \frac{wA_2\{1 - w^2 \exp[-(T+2T')/\tau]\}}{1 - w^4 \exp[-2(T+T')/\tau]}$$
(16)

The $\sigma_{xx}$ component of the conductivity tensor depends substantially on the probability for magnetic breakdown, because the charge carriers in a plane FS sheet moving on magnetic breakdown trajectories are subject to significantly less pronounced acceleration induced by the electric field, during their free path. For the same time period, the other $\sigma_{ij}$ components of the tensor are of the same order of magnitude as for $w = 0$.

The last term of eq. (16) decreases with an increase of $\tau$, so it cannot have strong influence on the asymptotic behaviour of the conductivity in strong magnetic field.

The contribution of the first term from eq. (16) to the component of the conductivity tensor is given by

$$\sigma_{xx} = -\frac{2e^3 H}{c(2\pi\hbar)^3}\int d\varepsilon \frac{\partial f_0(\varepsilon)}{\partial \varepsilon}\int dp_H \int_0^T dt v_x(t, p_H)$$
$$\{\int_0^t v_x(t', p_H)\exp(\frac{t'-t}{\tau})dt' - \frac{w^2 \exp[-(T+T')/\tau]}{1 + w^2 \exp[-(T+T')/\tau]}\int_{-T}^0 dt' v_x(t', p_H)\exp\frac{t'}{\tau}\}$$
(17)

It is negative, the two terms in the expression are of the same order of magnitude and are proportional to $\gamma$:

$$\sigma_{xx} = \sigma_0 \gamma \frac{1 - w^2 \exp(-(T+T')/\tau)}{1 + w^2 \exp(-(T+T')/\tau)}$$
(18)

As a consequence, the component of the resistivity tensor

$$\rho_{yy} = \frac{\sigma_{xx}}{\sigma_{xx}\sigma_{yy} - \sigma_{xy}\sigma_{yx}}$$
(19)

is proportional to $\gamma^{-1}$ and increases linearly with the magnetic field strength, when $\gamma << (1-w) << 1$.

In the immediate vicinity of Lifshic's phase transition, when $(1-w) \leq \gamma$, the linear increase of the resistance with the magnetic field strength is saturated. Quite the opposite, when the magnetic breakdown probability is very small, performing multiple recursive relations under $w <<1$, one gets:

$$\psi_1(\lambda_1+0) = \frac{A_1}{1-(1-w)\exp(-T/\tau)} \cong \frac{A_1}{w+T/\tau} \quad . \tag{20}$$

In such a way, the linear increase of the current resistance in the planes of the structure, can be witnessed in a wide range of magnetic fields ($\gamma << w(1-w)$).